# Optics experiments as a tool for developing critical thinking in physics education

**M Spodniaková Pfefferová[1*] and M Plesch[1,2]**

[1] Department of Physics, Faculty of Natural Sciences, Matej Bel University in Banska Bystrica, Banská Bystrica, Slovakia
[2] Institute of Physics, Slovak Academy of Sciences, Bratislava, Slovakia

[*]E-mail: miriam.spodniakova@umb.sk

**Abstract.** Experimental activities are an essential part of physics education. In addition to conveying scientific knowledge, they play a significant role in developing scientific literacy, inquiry skills, and critical thinking. In today's world, where students are exposed to vast amounts of information of varying quality, the ability to analyse, evaluate, and interpret information correctly has become increasingly important. This paper presents a series of physics experiments in the field of optics, specifically designed to foster critical thinking at various stages of the inquiry process. The topic of optics was chosen deliberately, as many optical phenomena occur naturally in everyday life, are familiar to students, and stimulate their curiosity. At the same time, they provide space for formulating hypotheses, designing experiments, interpreting data, and evaluating alternative explanations. Each experiment begins with a real-life problem situation that students are expected to explore and resolve through their own investigative work. The tasks are structured to encourage discussion, require argumentation, and promote reflection on both the process and the outcomes. The proposed experiments are suitable for students at both primary and secondary school levels and can be implemented in formal as well as non-formal educational settings. The aim of this contribution is to demonstrate how well-designed and pedagogically grounded experiments can not only enhance the understanding of physical concepts but also systematically develop critical thinking skills – one of the key competencies of 21st-century education.

## 1. Introduction

The main objective of the current school reform is to develop pupils' literacy skills. In the case of the educational area of Human and Nature (which consists of chemistry, biology, physics), it is the development of science literacy. Science literacy should be developed in such a way that pupils are able to identify the scientific aspects of many complex situations and to apply not only scientific knowledge, skills and attitudes, but also the principles of cognition in the natural sciences to these situations. Looking more closely at the stated aims, pupils should be able to compare, select, evaluate, justify and formulate explanations based on critical analysis of results and reasoning [1].

The goal of developing science literacy is not new. The development of science literacy has been addressed in various official documents in the past. According to the OECD, science literacy is defined as the ability to use scientific knowledge, identify issues and draw evidence-based conclusions to understand and make decisions about the natural world and the changes that have occurred in it as a result of human activity [2], [3]. The second part of the Human and Nature learning goal above, which speaks to the ability to work with information, critically evaluate it, and formulate conclusions, is more related to the development of critical thinking skills. It is the development of critical thinking skills that seems to be crucial today, given the amount of information we are confronted daily.

Several studies show that there is a positive and significant correlation between science literacy and critical thinking. A study [4] on 8th grade students showed that the higher the level of science literacy, the higher the level of critical thinking and problem-solving skills. Listiani et al. found a statistically significant correlation between science literacy and critical thinking in preservice teachers, confirming the importance of their joint development in the preparation of future teachers [5]. Moreover, Primasari et al. identified that science literacy contributes more to the development of critical thinking than student motivation alone [6]. Teaching methods such as project-based learning, assessment of science texts, or inquiry-based learning support the development of both domains simultaneously [7], [8].

From a didactic point of view, science literacy and critical thinking complement each other. Science literacy focuses on finding the best possible explanation of phenomena based on evidence and includes skills such as formulating hypotheses, experimenting, analysing data and constructing explanations. In contrast, critical thinking focuses on evaluating multiple alternative claims and selecting the most defensible view using both scientific and non-scientific (ethical, social) arguments. The two forms of thinking overlap, especially in the areas of metacognition, working with evidence, and logical reasoning [8].

To meet the stated goals, in terms of school physics, it is important to focus on the use of different methods of active cognition and systematic inquiry, so that pupils have as many opportunities as possible to learn selected methods of investigating physical phenomena in activities. In this paper we will focus on exploratory methods and their possibilities in developing pupils' critical thinking.

## 2. Inquiry and the development of critical thinking

As mentioned above, science literacy and critical thinking are interconnected. Critical thinking includes skills such as *interpretation, analysis, evaluation, inference, explanation and self-regulation* [9], [10]. These skills are not inborn but can be purposefully developed through appropriately chosen teaching methods.

In the teaching of science subjects, including physics, critical thinking is developed particularly in situations that require the formulation of hypotheses, the testing of theories, working with evidence and the ability to identify errors in reasoning. Experimental activities create natural conditions for the application of these processes.

If we look at inquiry-based activities, the stages of their implementation are generally: *setting the research problem and hypothesis, designing and planning the experiment, conducting the experiment, collecting experimental data, evaluating, analyzing and interpreting the experimental data, formulating generalized conclusions*. By comparing this information, we can conclude that, with an appropriately set way of conducting inquiry-based activities, we can also develop critical thinking in a targeted way. Pupils are encouraged to interpret a phenomenon they do not understand, then analyse the available information and formulate hypotheses which they themselves verify experimentally. In the next steps, they evaluate the data obtained, assess its relevance and reliability (evaluation) and formulate conclusions based on the evidence (inference, explanation). An important moment is also the reflection on the process and the correction of their own mistakes (self-regulation), which are essential elements of critical thinking.

Research confirms that inquiry-based learning (IBL) has the potential to develop critical thinking, especially when supported by discussion, open-ended questioning, and metacognitive reflection [11], [12]. Similarly, the PISA framework [3] emphasises the need to link scientific cognition with evidence evaluation and reasoning, which are key aspects of critical thinking in science contexts.

One possible strategy for developing critical thinking in inquiry-based activities is an assignment in which students identify a research problem based on a common life situation in which a physical phenomenon occurs. For example, the teacher poses a question to the pupils: '*Why do penguins congregate in large groups in freezing weather?*' The pupils' task is to formulate a research question based on observation and prior knowledge and then to test possible explanations using a simple experiment or model. In this way, pupils practise analysing a real situation, evaluating possible hypotheses and arguing their own conclusion, thereby actively engaging/developing the components of critical thinking.

Another appropriate method is to work with an incorrect or simplistic statement that contradicts the laws of physics but may seem intuitively correct. For example, the statement, "Heavier objects fall faster than lighter ones." The teacher gives the students the task of verifying this hypothesis experimentally. In designing the procedure, selecting appropriate objects and then implementing it, pupils examine, analyse and evaluate the evidence, leading them to reflect on their own thinking and correct misconceptions. According to research [13], [14], confronting error is one of the most effective ways to stimulate metacognitive processes and promote long-term cognitive change.

Critical thinking also develops in situations where pupils are working with multiple explanations for an observed phenomenon and must choose the one that is best supported by evidence. For example, when observing a 'broken' straw put in a glass of water, pupils may be offered a variety of possible ideas (e.g. 'the straw is bent', 'the water magnifies the image', 'the light changes direction'). The pupils' task is to critically evaluate each idea, discuss its validity and suggest a way of verification. This type of task encourages comparing alternative explanations, working with evidence, and the ability to justify their decisions based on logic and observation - key components of both science and critical thinking [15].

As mentioned above, the standards for the Human and Nature learning domain explicitly emphasize that students should be able to formulate research questions, generate hypotheses, design methods of experimental verification, and engage in critical analysis of results and reasoning. This corresponds closely with the earlier-mentioned inquiry-based activities, which have the potential to purposefully develop the cognitive aspects of critical thinking - from interpretation to analysis to reflection.

## 3. Using optical phenomena to develop critical thinking

Optics is an appropriate area of physics for conducting inquiry-based activities aimed at developing critical thinking. There are several reasons for this. First, it is a topic closely linked to the everyday experience of pupils – they encounter phenomena such as reflection, refraction, shadows, colours and rainbows in everyday life, but often do not understand them. At the same time, the visual nature of optical phenomena makes the field attractive and easy to observe, which encourages curiosity and motivation to explore.

Another advantage is the technical simplicity, many activities can be implemented with commonly available aids (mirrors, lenses, light sources, water, colour filters), which makes them suitable even for the conditions of a regular school. This article will present experiments in optics that are purposely designed to develop critical thinking at each stage of the inquiry-based activity - from the formulation of the research question, through the design and implementation of the experiment, to the interpretation of the data, formulating conclusions and defending them in a reasoned way.

To develop critical thinking skills in addition to inquiry skills in the proposed activities, a structure of worksheets has been prepared. A key element is the *initial motivational situation/question*, which plays an important motivational and cognitive role. This situation can either be based on a real-life situation (e.g. observation of a common optical phenomenon) or based on a physically incorrect statement that the pupils verify experimentally. The third type of task is situations where pupils are offered several possible explanations, and their task is to determine the correct one based on the evidence. Then there is an opportunity for *hypothesis formulation, information about the aids and space to record experimental data*. The subsequent phase, *analysing the data and interpreting the results*, plays a key role in the development of reasoning skills. Pupils not only record what they have observed, but learn to work with evidence, compare results with assumptions and draw conclusions that are supported by experimental data. The worksheet concludes with space for interpretation of results and possible discussion.

In the following section, we will present a few inquiry-based activities from Optics, focusing only on the description of the initial motivational part that was used in the activities and is crucial in terms of how the activity was implemented. The activities are mainly intended for secondary school, but their use in a modified form is also possible in primary school.

*3.1 How to become invisible*
The essence of the activity is an experiment with simple tools: vegetable oil and two beakers of different sizes or glass balls. The experiment is easy to carry out. After placing the smaller beaker (ball) into the larger one, the oil is slowly poured into it and the pupils watch the beaker/ball slowly disappear (Figures 1 and 2).

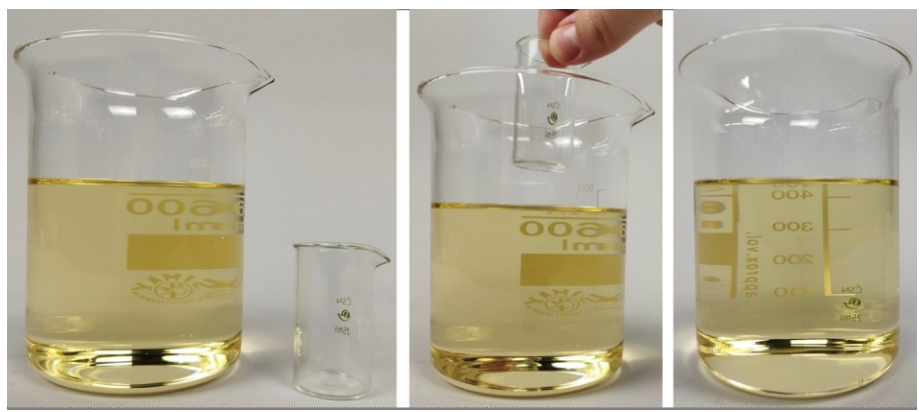

**Figure 1.** Disappearance of the beaker in vegetable oil
Author: Natália Demianová [16]

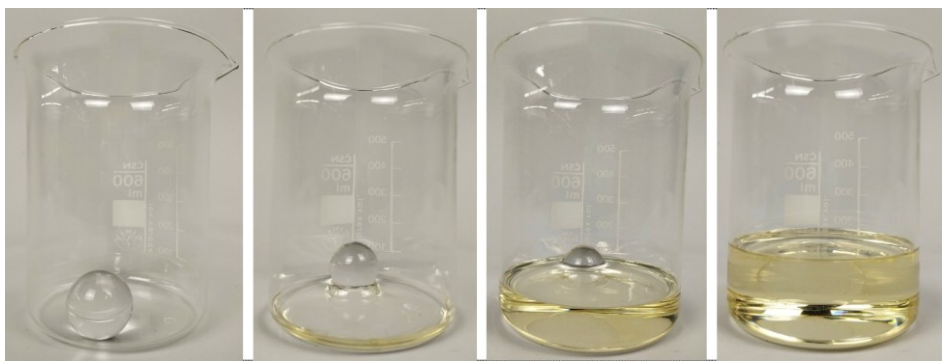

**Figure 2.** Disappearance of the glass small ball in vegetable oil
Author: Natália Demianová [16]

For this activity, two versions of the worksheets were created, which differed in the motivational introduction. While one worksheet contained in the introduction a description of a real situation from which it was possible to formulate a research problem/hypothesis, the other version contained in the introduction a physically incorrect statement, the correctness of which the pupils had to verify by experiment.

*1. version of the introductory part of the worksheet*
*Have you ever wondered **if it is possible to become invisible**? Maybe you've seen it in movies where the main character disappears thanks to a special cloak or technology. But the ability to "disappear" **doesn't just belong to the world of science fiction** – we encounter similar phenomena in the real world as well. In nature, for example, many animals **disguise themselves by blending in with their surroundings** – a chameleon changes colour, an insect pretends to be a leaf, hunter wears clothes that "hide" him in the forest.*
***But what about objects that become invisible without changing colour or shape?** Try to figure out how and why a glass bead/sphere can "disappear" in oil, and what this has to do with the light that surrounds it. Your task will be to figure out what determines whether we see an object or not*

*2. version of the introductory part of the worksheet*
*Imagine you are pouring oil into a glass at home and accidentally drop a small glass ball into it. After a moment of searching, you find that it is almost completely invisible - it is as if it has disappeared completely. During lunch you tell your older brother about it, and without hesitation he says: "That's because the oil dissolves the glass - in fact, it slowly disappears." Is that really so? **Try to verify this statement.***

Both versions of the worksheets share a common goal - the development of critical thinking in pupils, but with different didactical strategies. The first version works with open-ended questions framed by real-life examples. This approach leads the pupil to reflect on a phenomenon that is known but not understood, while at the same time creating the space to formulate their own research questions and hypotheses. Skills such as interpretation, situation analysis, question formation and problem identification are developed. The second version uses a situation from everyday life in which the pupil is confronted with a physically incorrect statement. This type of introduction purposefully creates cognitive conflict and activates the need for verification and reasoning. Thus, it promotes skills such as evaluating statements, drawing conclusions from evidence (inference) and metacognitive reflection on one's own ideas. Comparing the two approaches, it can be concluded that each of them develops a different type of thinking – the first one leads to divergent thinking (searching for questions, connections), while the second one promotes convergent thinking (verifying the truth of a particular statement).

For clarity, the key characteristics of both approaches are summarized in Table 1, which compares the results of their didactic analysis.

**Table 1.** Didactic analysis of both approaches

|  | 1. version of the introduction | 2. version of the introduction |
|---|---|---|
| **Didactic approach** | Open invitation to exploration | Confronting an incorrect hypothesis |
| **Critical thinking skills** | Interpretation, analysis, questioning | Evaluating claims, inference, correction |
| **Type of thinking** | Divergent (raising questions) | Convergent (verifying a specific statement) |
| **Emotional response** | Curiosity, fascination with a phenomenon | Doubt, need for verification, disagreement |
| **Starting point for discussion** | Multiple possible interpretations | One specific (incorrect) explanation |

It is clear from the table that these are different didactic strategies, but both can lead to the development of critical thinking. The choice of the appropriate strategy depends on the stated goal of the lesson, the target group.

Examples of other inquiry-based activities are accompanied by two versions of worksheets with different motivational introductory part, with the same goal - the development of critical thinking in pupils.

*3.2 Disappearing small balls*

The activity deals with the same physical phenomenon as the previous activity, except that hydrogel small balls are used instead of glass balls. The hydrogel balls can absorb a larger amount of water, which increases their volume several times. Pupils may be familiar with them as they are often used in various floral decorations (Figure 3).

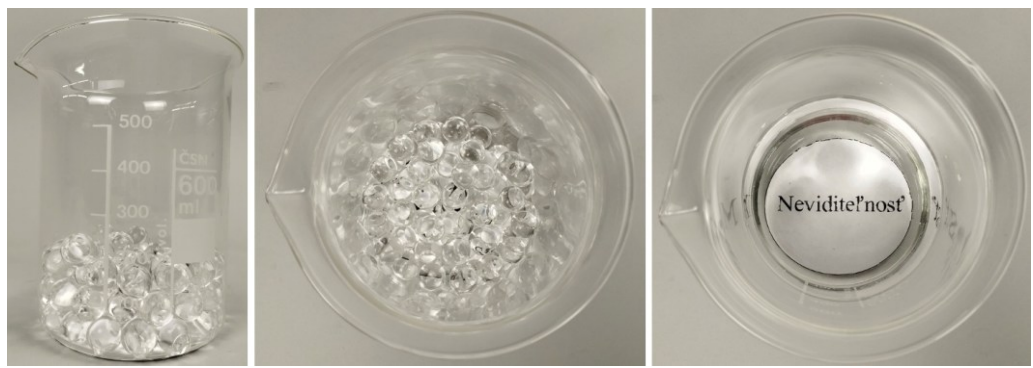

**Figure 3.** Hydrogel small balls a) without water and b) in water  
Author: Natália Demianová [16]

*1. version of the introductory part of the worksheet*
*Imagine you are preparing a transparent vase with flowers, and you want to decorate it with decorative gel beads. You put them in the vase, pour water over them - and in a moment they disappear. You can only see the water and the flowers, but when you reach into the vase with your hand, the beads are still there - you can feel them, but you can't see them. How is that possible?* ***Why are they visible when dry and "disappear" when water is poured over them?*** *What has changed – colour, shape, texture?*

*2. version of the introductory part of the worksheet*
*A short video of someone pouring gel balls into a container of water appeared on social media. After a while, they are no longer visible at all, and a comment appeared under the video: "**The balls dissolve quickly in the water, that's why you can't see them anymore.**" Sounds logical – or does it?* ***Is it really the case that they "dissolved"*** *or just stopped being visible for some other reason?*

At the end of the activity, regardless of the version of the worksheet used, pupils may be given a supplementary problem task in which they answer a question: "*What happens if we pour oil on the gel balls?*" In answering the question, pupils apply the knowledge they have obtained to a new situation, thus developing their ability to transfer knowledge and draw conclusions based on familiar phenomena. In terms of the development of critical thinking skills, such a task encourages the evaluation of hypotheses, analytical reasoning and the formulation of logically supported assumptions. If group discussion is used, pupils present their hypotheses to each other, compare them, argue them and form an opinion together, which also develops the ability to work together, to think argumentatively and to respect different points of view.

In the next activity we will again use water and oil, but this time the activity will focus on the topic of image formation by lenses.

*3.3 Can water change the course of events?*
In this activity, we will use a jar as a lens with which to create images of objects printed on paper. The essence of the activity is an experiment that appears quite often on social media, and the task of the activity is to explain the observed phenomenon. Again, there are two versions of the worksheets (Figure 4).

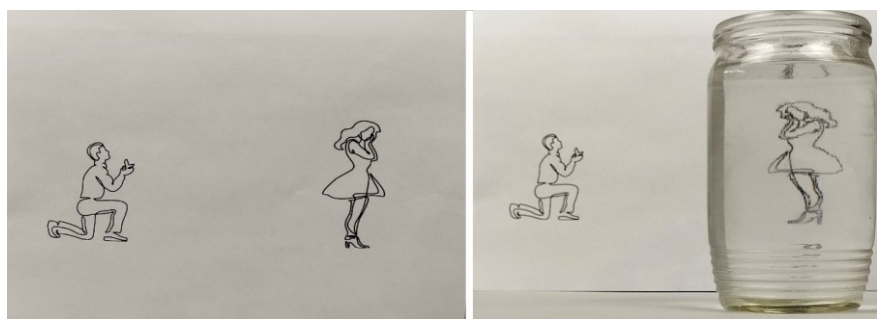
**Figure 4.** Representation of a young couple (a) without a jar, (b) using a jar
Author: Natália Demianová [16]

*1. version of the introductory part of the worksheet*
*Imagine a scene from a movie. A young man is standing in front of a girl, but her back is turned. The boy dares to ask her to marry him – but she doesn't seem to hear him at all. Everything changes*

*when a jar filled with water is placed between them. Suddenly, the girl seems to be looking at the boy. Can an ordinary jar filled with water change the course of events?*
*How is this possible? Why did the image change? And how does this relate to the shape and contents of the glass?* ***Your task will be to find out how an ordinary jam jar can create such an illusion.***

*2. version of the introductory part of the worksheet*
*Surely on social media have seen the video, where in the picture the boy asks the girl for her hand, but she stands back. In front of the girl is placed a jar with water and then the picture changes and the girl looks at the boy. One of the commenters below the photo said,* ***"The water in the glass acts as a mirror, so the girl's turn is reflected as if in a car's rear-view mirror." Is that really the case?***

Again, an additional question can be used at the end of the activity: 'What would happen if we used oil instead of water?' To find the answer to this question, pupils need to apply what they have already learnt to a new situation, not just what they have learnt in this activity. This will again lead to the development of critical thinking skills. The task requires pupils to apply their understanding of refractive index and light refraction in a different situation (application of new knowledge in new situation), identify which quantity is key to the formation of the inverted image (analysis), assess the probability of different hypotheses (evaluation). Finally, the learner must also self-reflect and critically assess his/her own hypothesis (self-regulation).

The use of the worksheets described above for inquiry-based activities is a good approach to developing pupils' critical thinking skills. The structure of the worksheets was designed to promote analytical thinking, inference, hypothesis evaluation and metacognition at each stage. Introductory motivational sections, based on either a real-life situation or an incorrect statement, create cognitive conflict and encourage pupils to independently verify and argue. The structured worksheets also allow for systematic recording of experimental data and conclusions, thus encouraging reflection and transfer of knowledge to new situations. In this context, worksheets are a suitable tool for integrating critical thinking into science education in line with the objectives of the school reform.

## 4. Conclusions

The inquiry-based activities described here were used in seminars and workshops for primary and secondary school pupils held at the Centre for Informal Physics Education at the authors' workplace during the 2024/2025 school year. Pupils carried out each activity using prepared worksheets. The primary objective was to test their usability, comprehensibility and didactic suitability from the point of view of the target group. Based on the feedback, the worksheets were subsequently revised and methodologically adapted.

Following this phase, it is planned to re-deploy the adapted worksheets in a workshop and seminar setting in the previously mentioned centre, in order to systematically test their potential to support the development of pupils' critical thinking at different stages of inquiry-based activities. Given the growing importance of critical thinking as one of the key competences for the 21st century, it is essential to pay increased attention to its development already in primary and secondary education. However, the effective implementation of these goals presupposes the availability of high-quality and didactically thought-out educational materials that will be of practical use to teachers in their daily teaching.

## Acknowledgement

Martin Plesch acknowledges additional support from the project no. 09I03-03-V04-00425 of the Research and Innovation Authority (Next Generation EU).

## References

[1] *Vzdelávacie oblasť Človek a príroda*. Retrieved June 12, 2025, from https://www.minedu.sk/data/ files/11815_clovek-a-priroda.pdf
[2] Lakatošová D and Veleg B 2015 *Tematická správa PISA 2006 Prírodovedná gramotnosť*. Retrieved June 15, 2025, from https://www2.nucem.sk/dl/3486/Tematická_správa_PISA_2006_-_prírodovedná_gramotnosť.pdf
[3] OECD 2023 *PISA 2025 Science Framework* (OECD Publishing) Retrieved June 15, 2025, from https://pisa-framework.oecd.org/science-2025/assets/docs/PISA_2025_Science_Framework.pdf
[4] Ridzal D A and Haswan 2023 *J. Pijar MIPA* **18** 1
[5] Listiani I *et al* 2022 *Al-Ishlah: Jurnal Pendidikan* **14** 721
[6] Primasari R *et al* 2020 *JPBI (Jurnal Pendidikan Biologi Indonesia)* **6** 273
[7] Vieira R M and Tenreiro-Vieira C 2016 *Int. J. Sci. Math. Educ.* **14** 659
[8] García-Carmona A 2023 *Sci. & Educ.* **34** 227
[9] Velmovská K *et al* 2021 *Aspekty rozvoja kritického myslenia vo vyučovaní fyziky* (Knižničné a edičné centrum FMFI UK, Bratislava)
[10] Facione P A 2015 *Critical Thinking: What It Is and Why It Counts* (Measured Reasons LLC, Hermosa Beach, CA)
[11] Hmelo-Silver C E *et al* 2007 *Educ. Psychol.* **42** 99
[12] Chin C and Chia L G 2004 *J. Biol. Educ.* **38** 69
[13] Kuhn D 1999 *Educ. Res.* **28** 16
[14] Osborne J 2014 *J. Sci. Teach. Educ.* **25** 177
[15] Zohar A and Dori Y J 2003 *J. Learn. Sci.* **12** 145
[16] Demianová N 2025 *Jednoduché experimenty z optiky* (Bachelor's thesis, Matej Bel University in Banská Bystrica, Faculty of Natural Sciences, Department of Physics)